\documentclass[prd,aps,floatfix,notitlepage,superscriptaddress,nofootinbib,twoco
lumn] { revtex4-1}

\usepackage{ucs}
\usepackage[utf8x]{inputenc}
\usepackage{amsmath}
\usepackage{amsfonts}
\usepackage{amssymb}
\usepackage{amsthm}
\usepackage[english]{babel}
\usepackage[T1]{fontenc}
\usepackage{graphicx}
\usepackage{nicefrac}

\usepackage[dvips]{hyperref}

\date{\today}

\begin{document}

\title{Glueballs from the Bethe-Salpeter equation}

\newcommand{\TUD}{Institut f\"ur Kernphysik, 
 TU Darmstadt, Theoriezentrum,
 Schlossgartenstra\ss e 2, 64289 Darmstadt, Germany}

\newcommand{\GU}{Institut f\"ur Theoretische Physik,
  Justus-Liebig-Universit\"at Gie\ss en,
  Heinrich-Buff-Ring 16,
  35392 Gie\ss en, Germany}

\author{Helios Sanchis-Alepuz}
\affiliation{\GU}
 
\author{Christian~S.~Fischer}
\affiliation{\GU}

\author{Christian Kellermann}
\noaffiliation

\author{Lorenz von Smekal}
\affiliation{\TUD}

 \begin{abstract}
 We formulate a framework to determine the mass of glueball 
 states of Landau gauge Yang-Mills theory in the continuum. To this end we 
 derive a Bethe-Salpeter equation for two gluon bound states including the 
 effects of Faddeev-Popov ghosts. We construct a suitable approximation scheme
 such that the interactions in the bound state equation match a 
 corresponding successful approximation of the Dyson-Schwinger
 equations for the Landau gauge ghost and gluon propagators. Based upon
 a recently obtained solution for the propagators in the complex momentum 
 plane we obtain results for the mass of the $0^{++}$ and $0^{-+}$ glueballs. 
 In the scalar channel we find a mass value in agreement with 
lattice gauge theory.
 \end{abstract}

\maketitle

\section{Introduction} \label{sec:Introduction}

The 'physical' spectrum of pure Yang-Mills theory is made out of 
glueballs \cite{Fritzsch:1975tx}. There is substantial evidence, 
both from lattice calculations (see \cite{Bowman:2007du,Maas:2011se} 
and Refs. therein) and results from Dyson-Schwinger equations (DSEs) 
for the Landau gauge gluon propagator
\cite{von Smekal:1997is,Alkofer:2003jj,complexglue} that transverse
gluons violate positivity and therefore cannot be part of the
asymptotic state space of the theory. Consequently, the first 
physical excitation of the Yang-Mills vacuum is the lowest lying
glueball state.  

It is an important task to determine the mass of this state. Indeed, 
the properties of glueballs have been investigated since their  
prediction in the middle of the 1970s \cite{Fritzsch:1975tx}. 
Today, the glueball masses in pure Yang-Mills theory are known rather 
accurately owing to high statistics lattice calculations 
\cite{Bali:1993fb,Morningstar:1999rf,Chen:2005mg}. Unquenched lattice 
calculations are also available, although there are considerable uncertainties 
in the determination of unquenched glueball masses 
\cite{Hart:2001fp,Bali:2000vr,Hart:2006ps,Richards:2010ck,Gregory:2012hu}. 
This is mainly due to severe problems with the signal to noise ratio, 
thus requiring large statistics. In principle, it is also not easy to
disentangle states with large glueball components from states dominated 
by other constituents such as quark-antiquark pairs. Naturally, this 
problem has a counterpart in the experiments: a glueball cannot be 
distinguished from a meson by quantum numbers and masses only. The 
determination of the decay channels of a given state is therefore vital
for its identification. Ongoing and new experiments such as BES III \cite{Asner:2008nq}
and PANDA \cite{Lutz:2009ff} have dedicated parts of their programs to the
identification of heavy glueballs in the charmonium region and beyond. 

Alternative theoretical frameworks such as Hamiltonian many body 
\cite{Szczepaniak:1995cw,Szczepaniak:2003mr,LlanesEstrada:2005jf,Bicudo:2006sd}
and strong coupling methods \cite{Pavel:2009nb}, 
potential approaches \cite{Brau:2004xw}, 
Wilson loop based calculations \cite{Kaidalov:1999de},
flux tube models \cite{Buisseret:2007de}, 
chiral Lagrangians \cite{He:2009sb,Janowski:2011gt,Eshraim:2012jv}, 
light front quantization \cite{Allen:1999kx}
and the AdS/QCD approach \cite{BoschiFilho:2012xr}
have shed some light on potential mass patterns and identifications of 
experimental states dominated by their glueball content. However, it seems
fair to state that a detailed understanding of glueball formation from the
underlying dynamics of Yang-Mills theory is still missing. In this paper we 
report on further steps towards such an understanding. 

Working in Landau gauge, we construct homogeneous Bethe-Salpeter equations 
(BSEs) for glueballs which take into account the dynamics of gluon 
and ghost propagation as well as their interactions with one another. This is 
detailed in Sections \ref{sec:Bound state equations} and \ref{sec:BS amplitudes}, 
where we also discuss the general form of Bethe-Salpeter vertices for 
any quantum number. The calculation is performed in  Euclidean space, 
which implies that the bound-state consitutents are 
probed for complex momenta. In the literature, exploratory BSE calculations 
using instantaneous approximations \cite{RaiChoudhury:1984ua} or extrapolations
of the propagators into the complex momentum plane \cite{Meyers:2012ka}
can be found. In this work we present first results for 
self-consistent and covariant BSE calculation of glueballs in Landau gauge 
using explicit solutions for the ghost and gluon propagator DSEs in the 
complex momentum plane \cite{complexglue}. These are summarized in Section 
\ref{sec:complexYM}.

\section{Bound state equations for glueballs} \label{sec:Bound state equations}

Our goal in this section is to provide a Bethe-Salpeter equation describing
a glueball made from two gluons that are solutions of the DSE for the gluon 
propagator. A similar concept has proven very successful in the context of 
mesons, where the BSE of a quark-antiquark pair is used in connection with 
the corresponding DSE for the quark propagator, see {\it e.g.}
\cite{Maris:2003vk,Fischer:2006ub,Bashir:2012fs} for reviews. A key 
property of this framework is consistency of the approximations made in the DSE 
and BSE. For mesons this implies to satisfy an axial Ward-Takahashi identity
thus implementing constraints due to chiral symmetry and its breaking. One way of 
devising such a truncation is to derive both the truncation of the 
DSE and the truncation of the BSE on common grounds using a two particle 
irreducible effective action (2PIEA). In the following we work along 
this strategy.
Since we are working in Landau gauge we need to take into account the Faddeev-Popov 
ghosts. Thus we need a generalization of the usual BSE scheme that allows 
for mixing of bound states of different fields. In the following we will give a 
derivation of a suitable set of bound state equations that provide the necessary 
couplings of bound state amplitudes with different field content 
\cite{Fukuda:1987su}. 

We consider the following 2PIEA,
\begin{widetext}
\begin{align} \label{eq:YMEA}
\Gamma[D,G]=&\frac{1}{2}\text{Tr}\ln\,D_0 D^{-1}+\frac{1}{2}\text{Tr}\, 
  D_0^{-1}D-\text{Tr}\ln\,G_0G^{-1}-\text{Tr}\,G_0^{-1}G+\Gamma_2[D,G],
\end{align}
\end{widetext}
where $D$ and $G$ are the gluon and ghost propagators. The interaction term is given diagrammatically by
%
\begin{align} \label{eq:YM2Lgrs}
\Gamma_2[D,G]= 
-\frac{1}{12}\,\parbox{2cm} {\includegraphics[width=2cm, keepaspectratio]{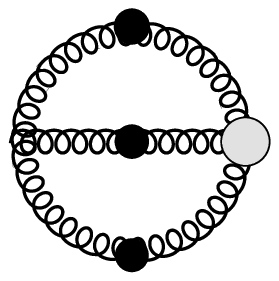}}
 +\frac{1}{2}\, \parbox{2cm}{\includegraphics[width=2cm, keepaspectratio]{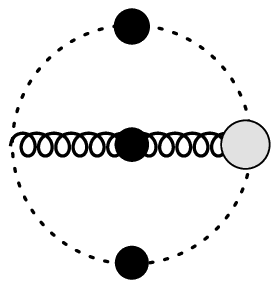}}.
\end{align}
%
Each term contains one bare and one dressed vertex, the latter being
represented by the shaded   
circles. The 2PIEA is already truncated, i.e. we have left out all 
diagrams including the four-gluon interaction. Furthermore, the dressed 
ghost-gluon and three-gluon vertices are assumed to be represented by 
suitable explicit expressions that capture the essence of the non-perturbative
interactions. Such Ansaetze have been employed successfully in the past 
\cite{Fischer:2006ub}; we come back to this point in Section \ref{sec:complexYM}. 

The corresponding Dyson-Schwinger equations for the ghost- and gluon propagators 
can be found by variation of the effective action with respect to a propagator, i.e.
\begin{align} 
 \frac{\delta\Gamma[D,G]}{\delta\,D}=&-D^{-1}+D_0^{-1}+\Sigma_D[D,G]=0 \label{eq:SDE1}\\
 \frac{\delta\Gamma[D,G]}{\delta\,G}=&\,\,2G^{-1}-2G_0^{-1}+\Sigma_G[D,G]=0 \label{eq:SDE2},
\end{align}
where we have $\Sigma_A=\frac{\delta\Gamma_2}{\delta\,A}$ with $A \in \{D,G\}$.
Diagrammatically, the resulting DSEs read
\begin{widetext}
    \begin{align} 
  \parbox{2cm}{\includegraphics[width=2cm, keepaspectratio]{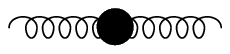}}^{-1} =& \quad
      \parbox{2cm}{\includegraphics[width=2cm, keepaspectratio]{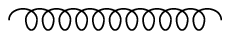}}^{-1} 
      -\frac{1}{2}\,\parbox{2.5cm}{\includegraphics[width=2.5cm, keepaspectratio]{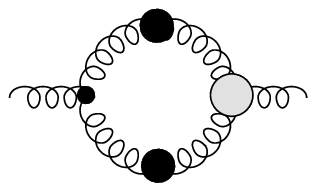}} 
      +\parbox{2.5cm}{\includegraphics[width=2.5cm, keepaspectratio]{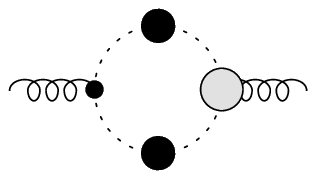}}\,,    
      \label{eq:redYM_System1} \\
 \parbox{2cm}{\includegraphics[width=2cm, keepaspectratio]{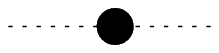}}^{-1} =& \quad
      \parbox{2cm}{\includegraphics[width=2cm, keepaspectratio]{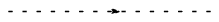}}^{-1}      -
      \raisebox{0.4cm}{
	\parbox{2.5cm}{\includegraphics[width=2.5cm, keepaspectratio]{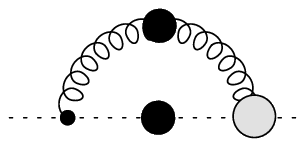}}
		      }\,.
	\label{eq:redYM_System2}
\end{align}
\end{widetext}
We now proceed along the lines of Ref.~\cite{Fukuda:1987su}. In the following 
we will use a shorthand notation omitting the space-time arguments and indicating 
primed arguments by primed functions. We denote the solutions of DSEs 
\eqref{eq:SDE1} and \eqref{eq:SDE2} by $\hat D$ and $\hat G$ and perform a
variation in two variables. Keeping only the linear terms we arrive at
\begin{widetext}
\begin{align}\label{eq:SDEexp2a}
 \frac{\delta\Gamma[D,G]}{\delta D}\bigg|_{{\hat D}+\delta_D, {\hat G}+\delta_G}
  \approx&\,\frac{\delta\Gamma[D,G]}{\delta D}\bigg|_{{\hat D}, {\hat G}}
  +\,\int d^4x^{\prime} d^4y^{\prime}\frac{\delta^2\Gamma[D,G]}{\delta D\delta D^{\prime}}
    \bigg|_{{\hat D}, {\hat G}} \delta_D^{\prime}+\,\int d^4x^{\prime} d^4y^{\prime}\frac{\delta^2\Gamma[D,G]}{\delta D\delta G^{\prime}}
    \bigg|_{{\hat D}, {\hat G}} \delta_G^{\prime}  \\
 \label{eq:SDEexp2b}
 \frac{\delta\Gamma[D,G]}{\delta G}\bigg|_{{\hat D}+\delta_D, {\hat G}+\delta_G}
  \approx&\,\frac{\delta\Gamma[D,G]}{\delta G}\bigg|_{{\hat D}, {\hat G}}
  +\,\int d^4x^{\prime} d^4y^{\prime}\frac{\delta^2\Gamma[D,G]}{\delta G\delta D^{\prime}}
    \bigg|_{{\hat D}, {\hat G}} \delta_D^{\prime}
    +\,\int d^4x^{\prime} d^4y^{\prime}\frac{\delta^2\Gamma[D,G]}{\delta G\delta G^{\prime}}
    \bigg|_{{\hat D}, {\hat G}} \delta_G^{\prime}. 
\end{align}
\end{widetext}

Using again the equations of motion we require for the solutions 
$\hat D$ and $\hat G$ to be stable that
\begin{align}
 \int d^4x^{\prime} d^4y^{\prime}&\frac{\delta^2\Gamma[D,G]}{\delta D\delta D^{\prime}}
    \bigg|_{{\hat D}, {\hat G}} \delta_D^{\prime}+\frac{\delta^2\Gamma[D,G]}{\delta D\delta G^{\prime}}
	\bigg|_{{\hat D}, {\hat G}} \delta_G^{\prime}=0 \label{eq:SDEexp3a} 
\end{align}
and
\begin{align}	
 \int d^4x^{\prime} d^4y^{\prime}&\frac{\delta^2\Gamma[D,G]}{\delta G\delta D^{\prime}}
    \bigg|_{{\hat D}, {\hat G}} \delta_D^{\prime}+\frac{\delta^2\Gamma[D,G]}{\delta G\delta G^{\prime}}
	\bigg|_{{\hat D}, {\hat G}} \delta_G^{\prime}=0. \label{eq:SDEexp3b}
\end{align}
Diagrammatically, the scattering kernels of the BSEs can be obtained by cutting 
a further line in the self energy diagrams with respect to the desired second 
constituent in the bound state. 
The variations $\delta^{\prime}_D, \delta^{\prime}_G$ are identified with the 
Bethe-Salpeter vertices $\chi_D$ and $\chi_G$.

We then find the following coupled system of BSEs for ghost and 
gluon bound states 
%
\begin{align}
\label{eq:YMBSE1}
\parbox{1.8cm}{\includegraphics[width=1.8cm, keepaspectratio]{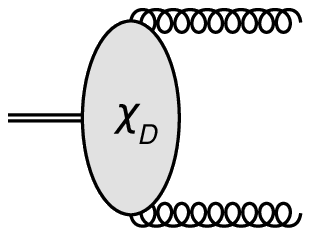}}=&\,\, 
  \parbox{2.5cm}{\includegraphics[width=2.5cm, keepaspectratio]{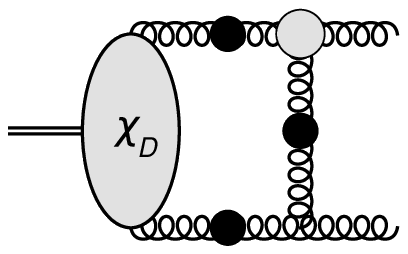}}-
  2\, \parbox{2.5cm}{\includegraphics[width=2.5cm, keepaspectratio]{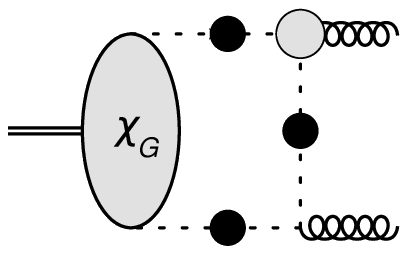}}
  +\updownarrow \,\\[3ex]
\label{eq:YMBSE2}
 \parbox{1.8cm}{\includegraphics[width=1.8cm, keepaspectratio]{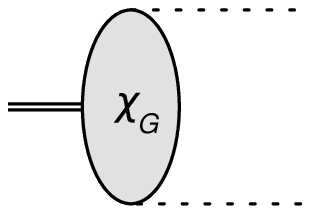}}=&\,
   \, \parbox{2.5cm}{\includegraphics[width=2.5cm, keepaspectratio]{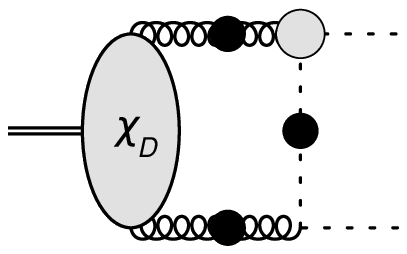}}+
 \parbox{2.5cm}{\includegraphics[width=2.5cm, keepaspectratio]{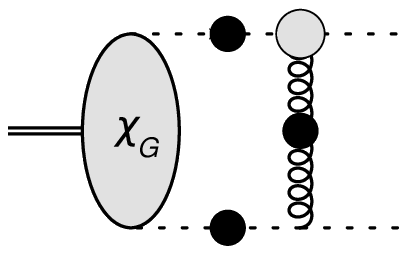}}
 +\updownarrow \, ,
\end{align}
%
where the arrow indicates symmetrization of the kernels with respect
to the dressed vertices. The resulting coupled system of two-body
equations serves to describe glueballs as bound states of either a
gluon or a ghost-antighost pair. The latter is necessary in Landau
gauge and represents contributions from the Faddeev-Popov determinant
to the glueball masses. We will later discuss  
the relative importance of both contributions in different channels. For now we
just emphasize that neither the ghosts nor the gluons are physical constituents
in the sense that they do not appear as propagating particles in the positive 
definite part of the asymptotic state space of 
QCD \cite{von Smekal:1997is,Alkofer:2003jj,complexglue}. 
Note, that there is no mixed gluon-ghost (or gluon-antighost) contribution 
to the glueball vertex. Such gluon-(anti)ghost bound states, if existent, 
may be members of a BRST quartet together with transverse gluons and, thus, 
part of the unphysical Hilbert space~\cite{Alkofer:2011pe}.

The above system of BSEs within Yang-Mills theory can be further 
generalized to full QCD by including quark contributions. Considering the 
corresponding effective action 
\begin{align} \label{eq:QCDEA}
\Gamma[D,G,S]=&\frac{1}{2}\text{Tr}\ln\,D_0D^{-1}+\frac{1}{2}\text{Tr}\,D_0^{-1}
D\nonumber \\ &-\text{Tr}\ln\,G_0G^{-1}-\text{Tr}\,G_0^{-1}G\nonumber \\ &
     -\text{Tr}\ln\,S_0S^{-1}-\text{Tr}\,S_0^{-1}S+\Gamma_2[D,G,S],
\end{align}
with diagrammatic representation
%
\begin{align} \label{eq:2Lgrs}
 \Gamma_2[D,G,S]= 
 -\frac{1}{12}\,\,\parbox{2cm}{\includegraphics[width=2cm, keepaspectratio]{Bilder/QCDeffAct1.eps}}
		 +\frac{1}{2}\,\,\parbox{2cm}{\includegraphics[width=2cm, keepaspectratio]{Bilder/QCDeffAct2.eps}}\\\nonumber+
		   \, \frac{1}{2}\,\,\parbox{2cm}{\includegraphics[width=2cm, keepaspectratio]{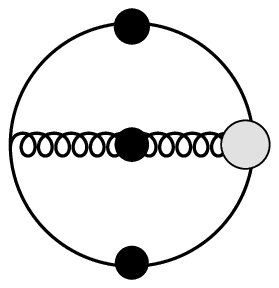}}\,,
\end{align}
%
we can apply the same derivation as before considering variations 
with respect to all types of propagators. 
We then find the full system of coupled bound state two-body equations
\begin{widetext}
\begin{align}
\label{eq:QCDBSE1}
\parbox{1.8cm}{\includegraphics[width=1.8cm, keepaspectratio]{Bilder/YMBSELHS1.eps}}=&\,
\parbox{2.5cm}{\includegraphics[width=2.5cm, keepaspectratio]{Bilder/YMBSERHS1a.eps}}-
   2\, \parbox{2.5cm}{\includegraphics[width=2.5cm, keepaspectratio]{Bilder/YMBSERHS1c.eps}}
   - 2 \, \parbox{2.5cm}{\includegraphics[width=2.5cm,keepaspectratio]{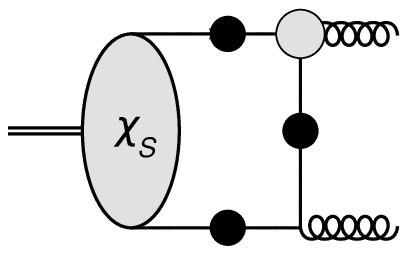}}+\updownarrow\\
\label{eq:QCDBSE2}
 \parbox{1.8cm}{\includegraphics[width=1.8cm, keepaspectratio]{Bilder/YMBSELHS2.eps}}=&\,
  \,\parbox{2.5cm}{\includegraphics[width=2.5cm, keepaspectratio]{Bilder/YMBSERHS2a.eps}}+
  \parbox{2.5cm}{\includegraphics[width=2.5cm,keepaspectratio]{Bilder/YMBSERHS2b.eps}}+\updownarrow\\
\label{eq:QCDBSE3}
 \parbox{1.8cm}{\includegraphics[width=1.8cm, keepaspectratio]{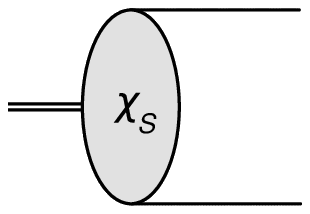}}=&\,
  \,\parbox{2.5cm}{\includegraphics[width=2.5cm, keepaspectratio]{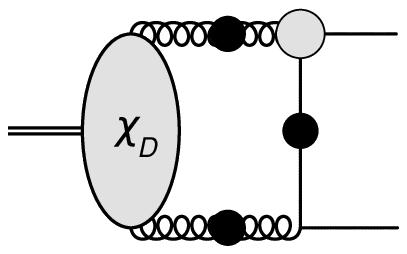}}+
  \parbox{2.5cm}{\includegraphics[width=2.5cm, keepaspectratio]{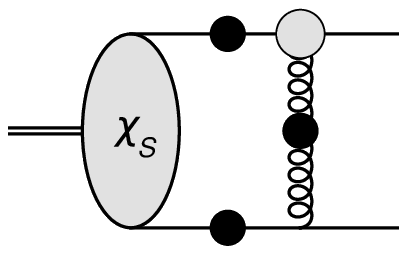}}+\updownarrow .
\end{align}
\end{widetext}

This set of BSEs describes mesons and glueballs in an approximation 
that can be seen as a generalized ladder truncation. Note, that 
the last diagram in \eqref{eq:QCDBSE1} and the first
of \eqref{eq:QCDBSE3} provide for glueball/meson mixing. Although 
in this work we will restrict our explicit calculations to pure
Yang-Mils theory we would like to add some comments on the influence 
of these terms onto the flavor singlet meson spectra. 

In the pseudoscalar channel these terms generate a contribution to 
$\eta$-$\eta^{\prime}$ splitting. In the framework of BSEs the conventional 
approach to this problem is to include beyond rainbow-ladder terms connected
with the axial anomaly in the quark-gluon interaction \cite{vonSmekal:1997dq,Bhagwat:2007ha,Alkofer:2008et}.  
Our framework provides the additional effect of a direct mixing of the flavor-singlet 
$\eta_0$-meson with the $J^{PC}=0^{-+}$ pseudoscalar glueball. Both effects
together affect the $\eta_0$-meson mass. Our set of BSEs provides the simplest 
means of consistently including such glueball/meson mixing into BSE calculations. 
However, since in lattice calculations the mass of the pure glue pseudoscalar 
glueball is found to be around $2.5\,\text{GeV}$ \cite{Chen:2005mg}, 
it is not clear how large these mixing effects might be.

Considerable mixing effects may be expected, however, in the scalar 
meson sector, where there seem to be more states than one can accommodate
in conventional quark anti-quark multiplets \cite{Nakamura:2010zzi}.
While some of these states can well be accounted for by large four-quark
components \cite{Jaffe:1976ig, Pelaez:2003dy,Heupel:2012ua}, others may very well be characterized
by a dominant glueball contribution. The set of Equations \eqref{eq:QCDBSE1} 
to \eqref{eq:QCDBSE3} may well provide a viable starting point for
sophisticated investigations of a realistic scalar meson spectrum.

In the following, we will focus on the coupled system of bound states 
for a pure gauge theory neglecting quarks (Eqs.~\eqref{eq:YMBSE1} and
\eqref{eq:YMBSE2}). To solve this system numerically, we need reliable
information on the nonperturbative  propagators of ghosts and gluons
as well as a general expression for the bound state vertices
$\chi_D$ and $\chi_G$. In the next section we will discuss the latter,  
providing suitable expressions for arbitrary quantum numbers of glueballs.

\section{Bound state vertices for glueballs} \label{sec:BS amplitudes}

We will now show how suitable bound state vertices $\chi_D$ and $\chi_G$ 
to be used in Eqs.~\eqref{eq:YMBSE1} and \eqref{eq:YMBSE2} can be constructed. 
We start our present discussion from some general observations. 

A bound state of two relativistic particles can be described by three 
quantum numbers: total spin $J$, parity $P$ and charge parity $C$. Furthermore 
there are only two characteristic momenta involved, conveniently chosen to be 
the total momentum $t_{\mu}$ and the relative momentum $r_{\mu}$.
These two vectors can be used to construct suitable vertices $\chi_D$ and 
$\chi_G$ for our bound state problem. The idea is to construct basic invariant 
vertices with correct parity and charge conjugation properties and supplement 
these with an appropriate tensor representing a given total spin. 

Let us first consider the vertex $\chi_D$ for the two-gluon bound state. A 
general scalar bound state vertex has to transform like a rank two Lorentz 
tensor for gluonic constituents, yet it has to be invariant under Lorentz 
transformations for it is meant to represent a scalar. Thus we are looking for 
a structure that transforms like
\begin{align} \label{eq:Amp1}
 {\Lambda^{\kappa}}_{\mu}\,{\Lambda^{\lambda}}_{\nu}\,T_{\kappa\lambda}=T_{\mu\nu}.
\end{align}
There exist only two tensor structures with $J=0$ that satisfy \eqref{eq:Amp1}, namely the metric tensor
and a combination of the totally antisymmetric tensor and characteristic momenta:
\begin{align}
\Gamma^{0^{++}}_{\mu\nu} = g_{\mu\nu}, 
\hspace*{2cm} \Gamma^{0^{-+}}_{\mu\nu} = \epsilon_{\kappa\lambda\mu\nu}r_{\kappa}t_{\lambda}
\label{basic}
\end{align}
Whereas the first represents a parity-even state, the second choice is odd under parity
transform. 

For the vertex $\chi_G$ composed of ghost fields the situation is trivial, 
since we are
looking for a term that couples to scalars and transforms like a scalar itself. 
The appropriate vertex is the identity in Lorentz space and has positive parity. This restricted 
choice has interesting implications as discussed below. 

In addition to the basic tensors in Eqs.~(\ref{basic}), representing
the Lorentz structure of the constituents, we need suitable tensors representing a given total momentum $J$
of the bound state.  A Lorentz tensor representing a massive field with  
total spin $J$ is required to have precisely $2J+1$ independent components to represent 
the possible spin polarizations. The construction of such 
tensors is known and a detailed treatment can be found e.g. in \cite{Zemach:1968zz}. We 
will repeat parts of the construction here in a slightly more explicit form focused 
directly on the construction of Bethe-Salpeter vertices. Consider first tensors 
$T_{a_1,...,a_J}$ in three-space of rank $J$. To represent angular momentum $J$ we 
require the tensor to be symmetric in all indices $T_{a_1...a_J}=T_{{\mathcal P}[a_1...a_J]}$
and traceless with respect to any pair of indices $\sum_m\,T_{...m...m...}=0$.
The first constraint leaves the tensor with $\frac{1}{2}\left(J^2+3J+2\right)$ 
independent components, while the second one imposes $\frac{1}{2}\left(J^2-J\right)$ 
further restrictions, thus leading to a tensor with $2J+1$ independent components, as
required.\footnote{It is also possible to construct tensors
  representing half-odd integer
 spin. Since we are dealing with Bethe-Salpeter equations of two particles 
in the same representation, so that the total angular momentum is
integer, we will not consider this possibility here but instead refer
the interested reader again to \cite{Zemach:1968zz}.} The construction of tensors in three-space is now easily transferred to four-tensors. 
If we require the tensor $T_{\mu_1...\mu_J}$ to be transverse to the total momentum of 
the particle in every index $t^{\nu}\,T_{...\nu...}=0$ and adopt the particles 
rest-frame, we see that all components with time-like indices vanish, leaving only 
components with space-like indices. So we are left with nothing else but the three-tensor 
considered before, which has $2J+1$ independent components. Thus we find the 
constraints for a Lorentz-tensor $T^{J}_{\mu_1...\mu_J}$ of rank $J$ to represent 
angular momentum $J$:
\begin{enumerate}
 \item $T$ is symmetric in all indices,
    \begin{align} \label{eq:LTsym}
     T^{J}_{\mu_1...\mu_J}=T^J_{{\mathcal P}[\mu_1...\mu_J]}.
    \end{align}
 \item $T$ is transverse to the total momentum of the particle in every index,
    \begin{align} \label{eq:LTtra}
     t^{\nu}\,T^{J}_{...\nu...}=0.
    \end{align}
 \item $T$ is traceless in every pair of indices in the rest-frame,
    \begin{align}
     T^{J,...\lambda...}_{...\lambda...}=0.
    \end{align}
\end{enumerate}
For the glueball masses we only need one such tensor from each multiplet. 
To construct such a tensor for angular momentum $J$ one can build the 
$J$-fold tensor product of a transverse projector that transforms like a vector 
and then subtract the traces with respect to every pair of indices. Starting with $J=1$, a suitable transverse four-vector can be obtained by contracting the   
transverse projector $\tau_{\mu\nu}$ (with respect to the total momentum $t$) 
and the relative momentum $r$,
\begin{align} \label{eq:Q_mu}
 Q_{\mu}= \tau_{\mu \nu} r^\nu = \left(g_{\mu\nu}-\frac{t_{\mu}\,t_{\nu}}{t^2}\right)\,r^{\nu}
 =\left(r_{\mu}-\frac{\left(r\cdot t\right)\,t_{\mu}}{t^2}\right).
\end{align}
With only one Lorentz index this transverse vector already gives a possible angular momentum tensor for $J=1$. For higher $J$ one builds symmetric $J$-fold tensor  products of \eqref{eq:Q_mu} by
\begin{align} \label{eq:rawTensors1}
 \tilde Q_{\mu_1...\mu_J}&=Q_{\mu_1}\times ...\times Q_{\mu_J}.
\end{align}
The next step is to remove the traces of these tensors with respect to every 
pair of indices. This can be achieved with the general formula
\begin{widetext}
\begin{align} \label{eq:tracelessT}
 T_{\mu_1...\mu_J}=&\tilde Q_{\mu_1...\mu_J}-
  (2J-1)^{-1}\,\sum_{P_{\mu_k}}\,\tau_{\mu_1\mu_2}\,{\tilde 
Q^{\kappa}}_{\;\;\kappa\mu_3...\mu_J}+
  &(2J-1)^{-1}\,(2J-3)^{-1}\,\sum_{P_{\mu_k}}\,\tau_{\mu_1\mu_2}\,\tau_{\mu_3\mu_4}\,{\tilde Q^{\kappa\lambda}}_{\;\;\;\;\kappa\lambda\mu_5...\mu_J}-
  \dots~,
\end{align}
\end{widetext}
where $\sum_P$ denotes the sum over all essentially different 
permutations of the indices.\footnote{This means that the sum 
has to be divided by appropriate combinatorial factors.} 
We furthermore define
\begin{align}
 f_2&=r^2-\frac{\left(r\cdot t\right)^2}{t^2}, \label{eq:f2}
\end{align}
and the tensors\footnote{For the convenience of the reader we have 
denoted the rank of the raw tensors as a superscript.}
\begin{align}
 B^{J,j}_{\mu_1...\mu_J}&=f_2^j\,\delta_{\{\mu_1\mu_2}...\delta_{\mu_{2j-1}\,\mu_{2j}}\,
  {\tilde Q}^{(J-2j)}_{\mu_{2j+1}...\mu_J\}},&2j<J,
  \label{eq:BTensor1}\\
 B^{J,j}_{\mu_1...\mu_J}&=f_2^{J/2}\,\delta_{\{\mu_1\mu_2}...\delta_{\mu_{J-1}\,\mu_{J}\}},
 &2j=J.
  \label{eq:BTensor2}
\end{align}
Using \eqref{eq:tracelessT} and \eqref{eq:rawTensors1} together 
with the above definition, we finally obtain the desired
total spin tensors in closed form as 
\begin{align}
 T_{\mu_1...\mu_J}&={\tilde Q}_{\mu_1...\mu_J}+\nonumber \\ &\sum^{2j\leq 
J}_{j=1}\,\left(-1\right)^j\,
  \frac{1}{j!\,2^j}\left(\prod_{k=1}^j\,2(J-k)+1\right)^{-1}\,B^{J,j}_{\mu_1...\mu_J} \label{eq:amQ}.
\end{align}
With these we have access to higher orbital angular momentum states built for glueballs with two gluon constituents (with $C=+1$) in the Lorentz singlet channels. For arbitrary even $J$ we can use,
\begin{align}
 \Gamma_{\mu\nu,\mu_1...\mu_J}^{J^{++}}(t^2,r^2,\theta)&=T_{\mu_1...\mu_J}\,A(t^2,r^2,\theta)\,g_{\mu\nu},
 \label{eq:amprad++even}\\
 \Gamma_{\mu\nu,\mu_1...\mu_J}^{J^{-+}}(t^2,r^2,\theta)&=T_{\mu_1...\mu_J}\,A(t^2,r^2,\theta)\,\left(r\cdot t\right)\,
  r^{\kappa}t^{\lambda}\,\epsilon_{\kappa\lambda\mu\nu},
 \label{eq:amprad-+even}
\end{align}
and if $J$ is odd,
\begin{align}
 \Gamma_{\mu\nu,\mu_1...\mu_J}^{J^{++}}(t^2,r^2,\theta)&=T_{\mu_1...\mu_J}\,A(t^2,r^2,\theta)\,\left(r\cdot t\right)\,
  r^{\kappa}t^{\lambda}\,\epsilon_{\kappa\lambda\mu\nu},
 \label{eq:amprad++odd}\\
 \Gamma_{\mu\nu,\mu_1...\mu_J}^{J^{-+}}(t^2,r^2,\theta)&=T_{\mu_1...\mu_J}\,A(t^2,r^2,\theta)\,g_{\mu\nu}.
 \label{eq:amprad-+odd}
\end{align}
Here we have introduced scalar functions $A(t^2,r^2,\theta)$, 
which are even under inversion of the angle $\theta$ between $r$ and $t$.
The additional factors $(r\cdot t)$ ensure the correct behavior of the bound 
state vertices under charge parity transformations, which result in a simple
flip of the sign of the relative momentum $r_{\mu}$ in our framework.

The corresponding vertices for glueballs with a ghost anti-ghost pair as 
constituents are constructed along the same lines. For arbitrary positive parity
even $J$ we can use
\begin{align}
 \Gamma_{\mu_1...\mu_J}^{J^{++}}(t^2,r^2,\theta)&=T_{\mu_1...\mu_J}\,A(t^2,r^2,\theta),
 \label{eq:ghost++even}
\end{align}
and if $J$ is odd the negative parity states are obtained from
\begin{align}
 \Gamma_{\mu_1...\mu_J}^{J^{-+}}(t^2,r^2,\theta)&=T_{\mu_1...\mu_J}\,A(t^2,r^2,\theta).
 \label{eq:ghost-+odd}
\end{align}
Note that the parity $P$ of the angular momentum tensors $T_{\mu_1...\mu_J}$ is given by $P=(-1)^J$. Thus the parity
of the even (odd) $J$ tensors is positive (negative). 
Hence there are neither contributions from ghost anti-ghost
pairs to glueballs for even $J$ and with quantum numbers $J^{-+}$, nor
for odd $J$ with $J^{++}$. For the gluonic vertices 
the natural parity of the total spin tensors can be supplemented by the odd
parity basis element given in \eqref{basic}, thus changing the overall parity 
of the tensor representation. This is not possible for the ghost vertices. 
Consequently, we only find the restricted set \eqref{eq:ghost++even} and 
\eqref{eq:ghost-+odd} of possible quantum numbers for glueball states with 
ghost contributions. 

Another potential restriction for the contribution of gluonic vertices
to glueballs has been frequently discussed in the literature in the context 
of model building \cite{Barnes:1981ac}: if the gluonic constituents were
massless and on-shell, \textit{Yang's theorem} \cite{Yang:1950rg} would
restrict the number of allowed quantum numbers drastically. In our framework
this constraint appears to be almost irrelevant. The non-perturbative 
gluonic constituents that appear in the BSE are neither on-shell nor 
massless. Instead they acquire a dynamically generated mass, as discussed 
in more detail in the next section.\footnote{Note that in the context of this 
discussion of Yang's theorem it is irrelevant whether the mass generation 
mechanism leads to an infrared vanishing gluon propagator (`scaling') or 
an infrared finite propagator (`decoupling'). Furthermore, 'mass generation' 
in this context does not mean that the gluon propagator acquires a pole
at time-like momenta, but merely that the zero-momentum pole of the free 
propagator disappears due to interactions.} Thus, gluonic contributions
in Yang-forbidden channels may be suppressed but certainly not forbidden.
These channels are the $1^{++}$ and all odd $J^{-+}$-channels. Since in the
$1^{++}$-channel ghost contributions are absent as well, the potentially 
suppressed gluonic contributions may lead to an `unnaturally' large glueball
mass in this channel. This is indeed observed in lattice 
calculations \cite{Morningstar:1999rf,Chen:2005mg}. In the odd
$J^{-+}$-channels, however, ghost contributions are allowed. If the suppression
of the gluonic contributions were strong, these states could be termed
`ghostballs'. We will study such states in future work. 

In addition, there may be another basic vertices for the gluons in the spin 2 channel, traceless symmetric tensors. For example, Landau constructed one for QED with $J^{PC}=2^{++}$ \cite{Landau:1948sa}. The explicit form of this basic $J=2$  tensor with our notations would read
\begin{widetext}
\begin{align} \label{eq:basic_spin2}
 \nonumber
 \Gamma^{2^{++}}_{\mu\nu;\mu_1\mu_2}=& \, t^4\Bigl(-\frac{1}{3}g_{\mu\nu}g_{\mu_1\mu_2}+\frac{1}{2}g_{\mu\mu_1}g_{\nu\mu_2}+\frac{1}{2}g_{\mu\mu_2}g_{\nu\mu_1}\Bigr) \\
 \nonumber
 & \,\, + t^2\Bigl(\frac{1}{3}g_{\mu\nu}t_{\mu_1}t_{\mu_2}-\frac{1}{2}g_{\mu\mu_1}t_{\nu}t_{\mu_2}-\frac{1}{2}g_{\mu\mu_2}t_{\nu}t_{\mu_1}                   
  - \frac{1}{2}g_{\nu\mu_1}t_{\mu}t_{\mu_2}-\frac{1}{2}g_{\nu\mu_2}t_{\mu}t_{\mu_1}+\frac{1}{3}g_{\mu_1\mu_2}t_{\mu}t_{\nu}\Bigr)                          \\
 & \,\, + \frac{2}{3}t_{\mu}t_{\nu}t_{\mu_1}t_{\mu_2},
 \end{align}
 \end{widetext}
where the pair $(\mu,\nu)$ denotes the Lorentz indices of the gluon constituents and 
$(\mu_1,\mu_2)$ the one of the bound state. With such a tensor vertices for a 
given set of quantum numbers $J^{PC}$ can be constructed in a similar way as from the two singlet tensors in Eqs.~(\ref{basic}). 

Having discussed the derivation of the bound state equations we will use to 
describe glueballs and the form of the necessary bound state vertices, we 
will now turn to the numerical part of our investigation. We will find that 
it is necessary to solve the system of coupled DSEs of ghost and gluon fields 
for complex momenta in order to use the resulting propagators in our calculation 
of glueballs.

\section{The Yang-Mills system in the complex plane} \label{sec:complexYM}
\begin{figure*}[htbp]

\centerline{\includegraphics[width=0.9\columnwidth]{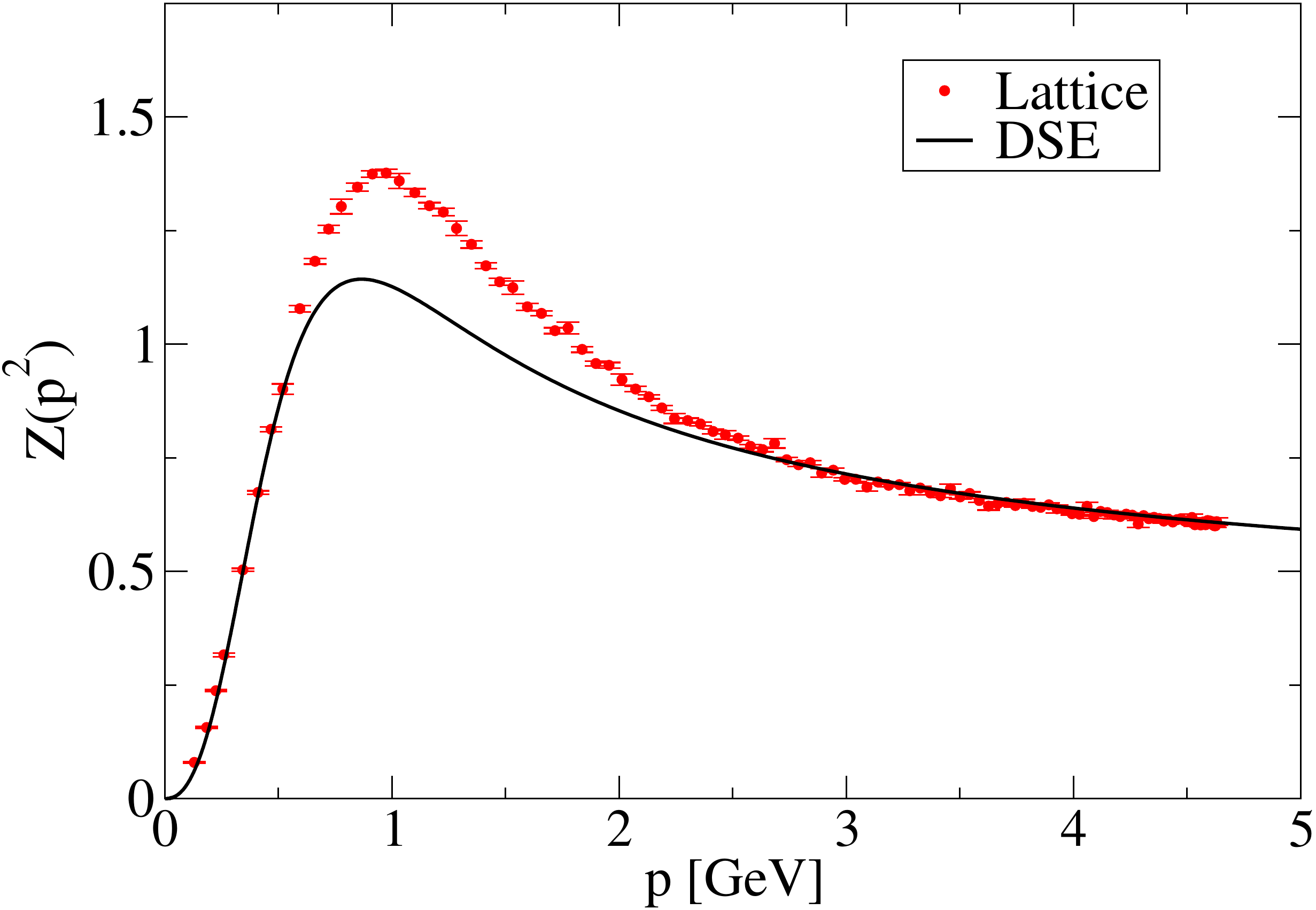}\hfill   
\includegraphics[width=0.9\columnwidth]{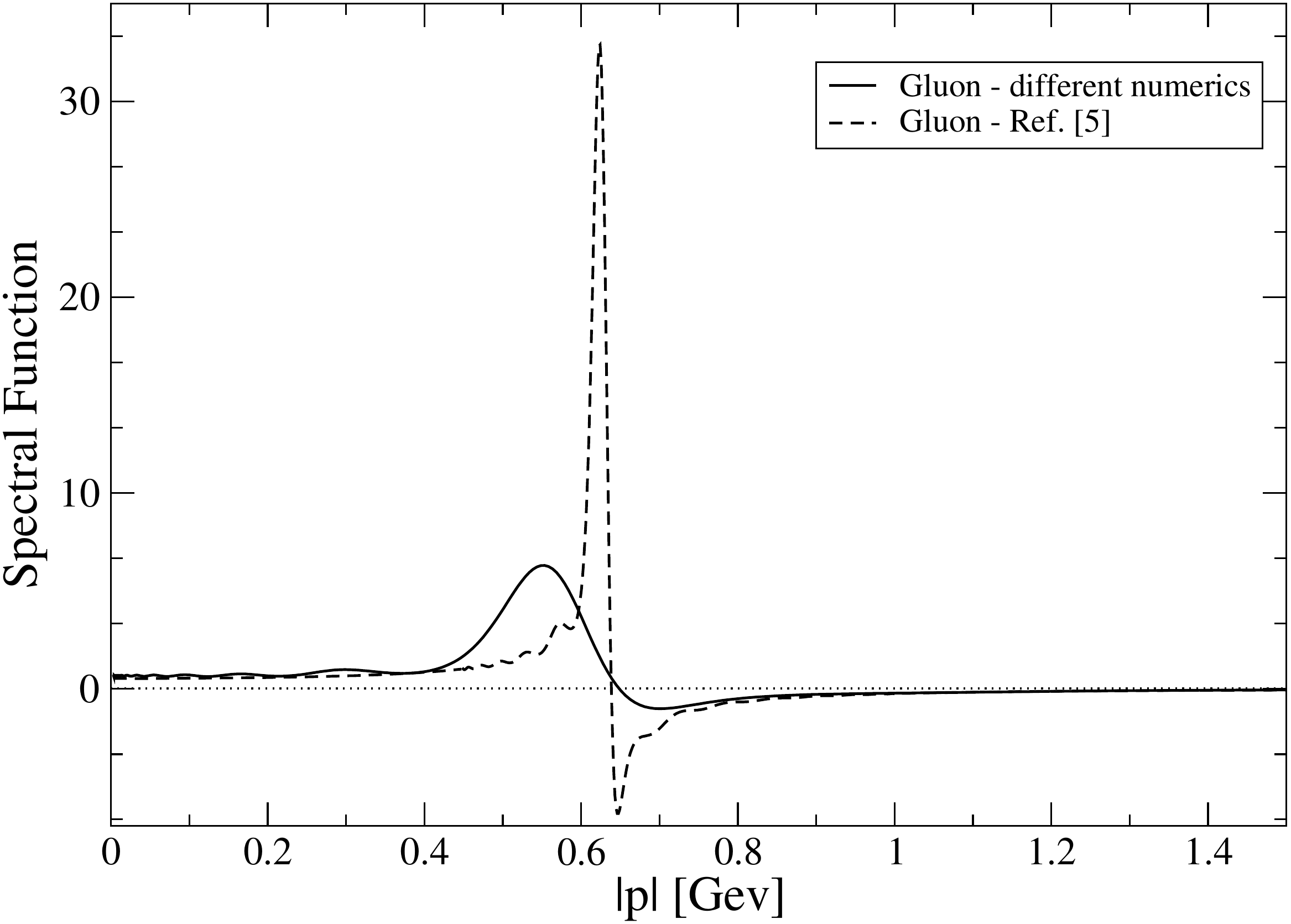}}
\caption{Left: Results for the gluon dressing function $Z(p^2)$
  from the DSEs \cite{Fischer:2008uz} for real and space-like momenta, compared
  with lattice calculations \cite{sternbeck06}. Right: Results for the gluon
  spectral function from DSEs for time-like momenta. Shown is the result from 
  Ref.~\cite{complexglue} together with result obtained in the same truncation 
  scheme but with slightly different numerics, see main text for further 
  explanations.}
\label{fig:real}
\centerline{\includegraphics[width=0.9\columnwidth]{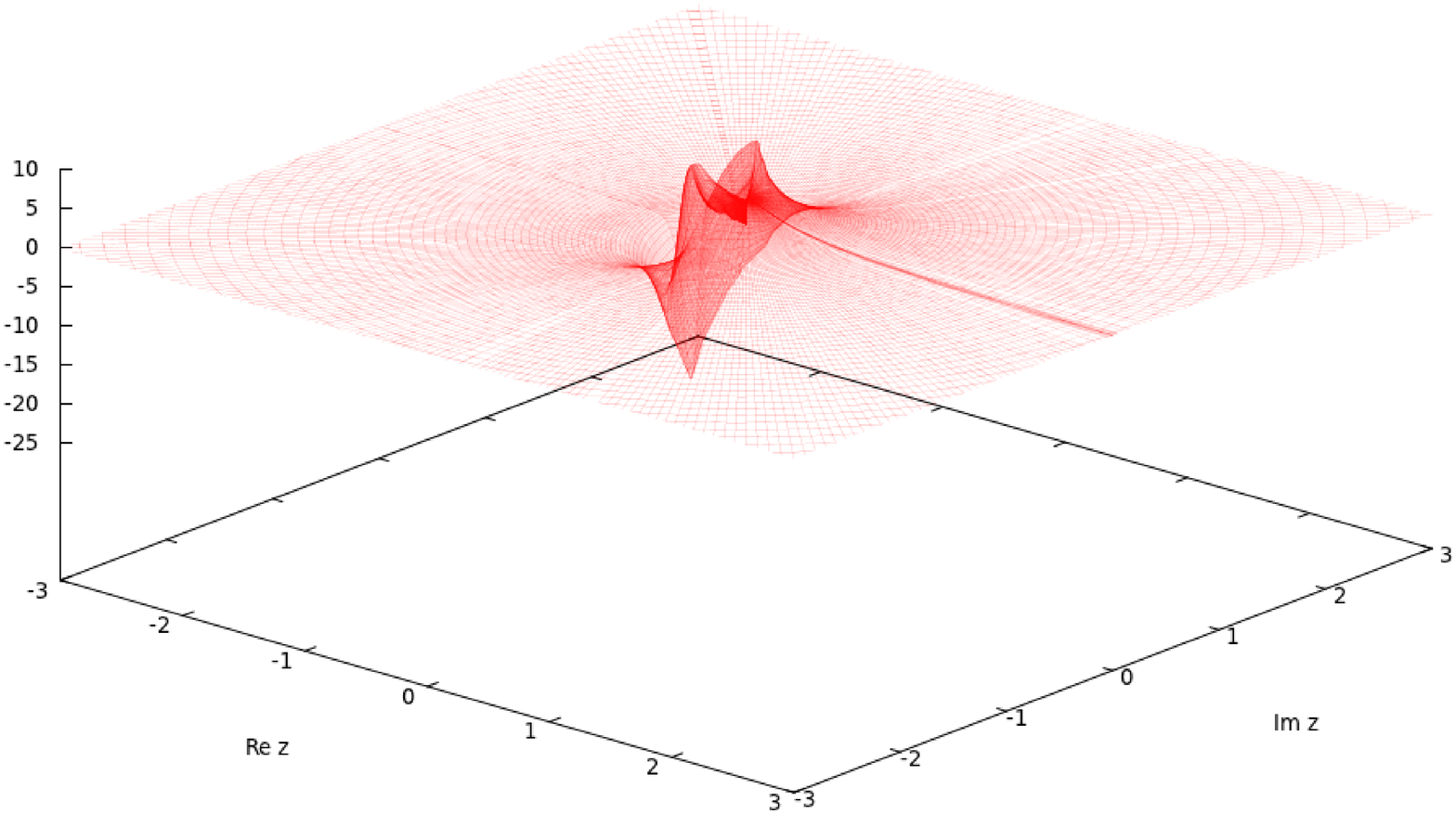}}
\caption{Real part of the gluon propagator function $D(p^2) = Z(p^2)/p^2$ in 
 the complex $z=p^2$ plane \cite{complexglue}.}
\label{fig:complex}

\end{figure*}

We solve the coupled system of bound state equations, \eqref{eq:QCDBSE1} 
and \eqref{eq:QCDBSE2} in pure Yang-Mills theory in Euclidean momentum space. 
In the rest frame of the glueball, its total momentum is then given by
$(0,0,0,im_B)$ with $m_B$ is the bound-state mass. 
Without loss of generality, the total momenta can be shared equally between the
two constituents. Their momenta are then given by $r_\pm = \left(r \pm 
t\right)/2$, with
relative momentum $r$ between the constituents. It is then clear that the 
internal propagator lines 
in BSEs are given by solutions of the DSEs for complex momenta.

These can be obtained from the corresponding coupled set of Dyson-Schwinger 
equations. In Landau gauge the ghost propagator $D_G(p^2)$ and the gluon 
propagator $D_{\mu \nu}(p^2)$ are given by
\begin{align}
D_{\mu \nu}(p^2) =& \left(\delta_{\mu \nu} - \frac{p_\mu p_\nu}{p^2}\right) 
\frac{Z(p^2)}{p^2}\,,\\
D_G(p^2) =& -\frac{G(p^2)}{p^2}\,,
\end{align}
where the diagonal color structure has been omitted for brevity. Note that the 
gluon is transverse also non-perturbatively, thus spurious glueball states due 
to longitudinal modes, as present in some potential models~\cite{Brau:2004xw}, 
are naturally avoided. The coupled system of DSE (omitting two-loop diagrams) 
has been displayed diagrammatically in Eqs.(\ref{eq:redYM_System1})-(\ref{eq:redYM_System2}). 
This system of equations has been considered frequently in the past years. 
It has been solved analytically in the deep infrared, where exact solutions 
without any truncations are possible \cite{Lerche:2002ep,Alkofer:2004it}. 
Two qualitatively different solutions have been found named `scaling' and 
`decoupling' \cite{Fischer:2008uz,Aguilar:2008xm}. Whereas 
the scaling solution consists of infrared
power laws for all Green's functions with an infrared vanishing gluon propagator
and an infrared divergent ghost, the decoupling solution is characterized by an
infrared finite gluon propagator and a finite ghost dressing function. Current lattice
calculations on large volumes clearly favor the decoupling type of solutions
\cite{Cucchieri:2008fc}; there is, however, an ongoing discussion on potentially
significant effects from different gauge fixing strategies in the deep infrared 
\cite{vonSmekal:2008ws,Sternbeck:2008mv,Cucchieri:2009zt,Maas:2009se,Sternbeck:2012mf,Dudal:2014rxa}.
In this work we concentrate on decoupling type of solutions, which have been associated
with a dynamically generated `gluon mass'.\footnote{Of course, this `mass' is not 
to be identified with the mass of a physical particle. The analytic structure of the 
gluon propagator is clearly different from a simple mass pole \cite{complexglue}.} 
As already mentioned above, these still maintain transversality.
At finite momenta, the equations have to be solved numerically and approximations for the 
dressed vertices need to be introduced. A suitable truncation scheme has been
introduced in Ref.~\cite{Fischer:2002hna} and improved in Ref.~\cite{Fischer:2008uz}.
It involves educated ansaetze for the ghost-gluon and three-gluon vertex and neglects
the effects of the four-gluon interaction completely. The resulting 
modified system of equations
has exactly the structure of the system \eqref{eq:SDE1} and \eqref{eq:SDE2} 
obtained from the variation of the 2PIEA \eqref{eq:YMEA} in 
Section \ref{sec:Bound state equations}. For real momenta, we show the 
numerical solution for the gluon dressing function in Fig.~\ref{fig:real}.
As can be seen from the comparison with the lattice results \cite{sternbeck06},
there is very good agreement in the infrared and ultraviolet momentum region,
whereas in the mid-momentum region one observes quantitative
deviations. These deviations are certainly in parts due to the
neglected four-gluon interactions in the DSEs (\ref{eq:redYM_System1})-(\ref{eq:redYM_System2}). 
They can be compensated, however, by suitably
optimizing the input used for the dressed three-gluon vertex by
simultaneously solving its own DSE together with those for the propagators
\cite{Huber:2012kd}.  

The system of DSEs \eqref{eq:redYM_System1}, \eqref{eq:redYM_System2} for the 
ghost and gluon propagator has been solved in the complex $p^2$ plane recently, 
see Ref.~\cite{complexglue} for details. 
As explained above, this complex solution constitutes a vital input into the 
corresponding Bethe-Salpeter equation for the glueballs and is used in the 
following. In this respect it is important to note that the analytic structure 
of the gluon and ghost propagators as obtained in Ref.~\cite{complexglue}
shows branch cuts along the time-like momentum axis, i.e.~for
negative invariant momentum squared, but no
singularities away from the real axis in the complex $p^2$ plane. For the real part of the gluon propagator this can be 
seen in Fig.~\ref{fig:complex}, corresponding plots for the imaginary part and 
the ghost dressing function can be found in Ref.~\cite{complexglue}. This behavior 
greatly helps in the numerical treatment of the BSE. Along the
time-like axis of negative $p^2$ one can extract the gluon 
spectral function, which is shown in the right plot of Fig.~\ref{fig:real}.
We have plotted the result from Ref.~\cite{complexglue} together with a 
corresponding result obtained from an improved numerical method. The corresponding 
results are very similar except on a narrow region around the cut on the negative 
$p^2$-axis. As a result, one obtains a considerable smoother spectral function,
as can be seen in Fig.~\ref{fig:real}. We use this improved result in the present work.

Finally, a comment on the scale is in order. This is fixed by comparison with 
the lattice results for the gluon propagator and remains fixed, i.e.~it is not 
adjusted again in the glueball calculations. Thus, in principle, we obtain absolute 
values for the glueball masses.   

\section{Lowest lying glueball masses} \label{sec:Masses}

We have solved the BSEs for a glueball in the scalar and pseudoscalar channel 
using the bound state equations \eqref{eq:YMBSE1} and \eqref{eq:YMBSE2} together
with the vertices \eqref{eq:amprad++even} to \eqref{eq:amprad-+odd}.
For the propagators of ghosts and gluons we use the numerical results
discussed in the last section. 

Bound state vertices are not primitively divergent vertices and therefore they 
generically go to zero like power laws for large momenta \cite{Aoki:1990aq}.
In contrast to the behavior of meson BSEs, however, \eqref{eq:YMBSE1} and 
\eqref{eq:YMBSE2} also admit solutions with a logarithmic behavior in the UV.
(Details on the asymptotic behavior of glueball BSEs together with an explicit 
analytic analysis will be given elsewhere.) These solutions do not correspond 
to bound states. In order to guide the iterative numerical procedure to the correct 
bound state solution it turns out to be sufficient to introduce an additional 
Pauli-Villars term into the purely gluonic diagram of \eqref{eq:YMBSE1} that 
depends on the momentum of the exchanged gluon propagator. Namely, we replace 
$Z(k^2)\rightarrow Z(k^2)\left(1+k^2/\Lambda^2_{PV}\right)^{-1}$, 
with $k^2$ the momentum of the exchanged gluon and $\Lambda_{PV}$ a cutoff 
scale. By inspection of the Bethe-Salpeter vertex functions we have verified, 
that such a term does not simply modify the logarithmic solutions of the BSE
above the scale $\Lambda^2_{PV}$ but indeed drives the equation to a different
and well-behaved solution. We find that the resulting glueball masses are 
insensitive to all values $\Lambda^2_{PV} > 100~$GeV$^2$ of the scale that 
we have probed.

We present our results in 
Table \ref{tab:glueball_masses} together with corresponding ones from 
lattice gauge theory, the Hamiltonian approach and Regge theory. Additionally, 
we compare to a rather recent calculation in a non-relativistic constituent model. 
\begin{table}[t]
 \begin{center}
 \begin{tabular}{|c|rl|rl|rl|rl|}
  \hline
                 & \multicolumn{8}{c|}{masses (GeV)}\\
  \hline
  $J^{PC}$       & \multicolumn{2}{c|}{lattice}    & \multicolumn{2}{c|}{Hamiltonian/} & \multicolumn{2}{c|}{constituent} & \multicolumn{2}{c|}{this work} 	\\
                 &      &                          & \multicolumn{2}{c|}{Regge theory} & \multicolumn{2}{c|}{models}      &&								\\
  \hline
  $0^{++}$       & 1.71 (5)(8) &\cite{Chen:2005mg}        & 1.98 
&\cite{Szczepaniak:2003mr}   & 1.71 &\cite{Buisseret:2009yv}    & 1.64 	&	
					\\
		 		 & 1.73 (5)(8) &\cite{Morningstar:1999rf} & 1.58 
&\cite{Kaidalov:1999de}      & 1.86 &\cite{Brau:2004xw}         &			
&						\\
  \hline
  $0^{-+}$       & 2.56 (4)(1) &\cite{Chen:2005mg}        & 2.22 
&\cite{Szczepaniak:2003mr}   & 2.61 &\cite{Buisseret:2009yv}    & 4.53 	&	
					\\
		 		 & 2.59 (4)(13) &\cite{Morningstar:1999rf} & 
2.56 &\cite{Kaidalov:1999de}      & 2.49 &\cite{Brau:2004xw}         &		
	&						\\
 \hline
 \end{tabular}
 \end{center}
 \caption[Glueball mass spectrum for $J=0$.]
    {Scalar and pseudoscalar glueball masses (in GeV) from various studies. We 
quote the Model B data from Ref.~\cite{Brau:2004xw}.}
 \label{tab:glueball_masses}
\end{table}

Comparing with the lattice results, we find that the state with quantum numbers 
$0^{++}$ is well reproduced on the five percent level. 
Compared with the lattice, the good agreement of our result 
for the lowest lying scalar glueball is remarkable, though probably not 
surprising. As explained above, our truncation scheme for the ghost/gluon DSEs produces
solutions which are point-wise similar to the lattice results in the low and high
momentum region and display a twenty percent difference for momenta around 1 GeV.
Thus the overall quality of the truncation scheme is well below the twenty percent 
range and thus in agreement with our findings for the scalar glueball mass. The 
remaining deficiencies in our truncation scheme are in the details of the three-gluon
and the missing four-gluon interactions. 

In contrast, the mass of the pseudoscalar glueball is much higher than that 
predicted by lattice calculations as well as by other approaches. 
As discussed in Section \ref{sec:BS amplitudes}
there are no ghost contributions in these channels, leaving a greatly reduced 
BSE with only one gluonic diagram to be solved. This diagram 
is, in turn, largely dominated by the three-gluon vertex, both directly and 
via the solution of the DSE for the gluon propagator. Since the ansatz used 
here was devised in the context of the study of gluon and ghost DSEs for real 
momenta, it is conceivable that the behavior of this ansatz in the complex plane 
affects significantly the glueball spectrum in particular for states with no
ghost-antighost content. The study of the connection between the details of the
non-perturbative gluon self-interactions and their impact on glueball masses 
will be the subject of future work.

\section{Summary} \label{sec:Summary}

In this paper we have presented a framework that allows to calculate glueball 
properties from the dynamics of Landau gauge Yang-Mills theory. We have constructed 
a set of bound state equations that includes both ghosts and gluons degrees of freedom 
thus taking into account also the effects of the Faddeev-Popov determinant. This set of 
equations allows for mixing of bound state contributions from different species 
of particles and is readily generalized to full QCD, including quarks. It thus naturally 
incorporates meson/glueball mixing. Furthermore we have presented suitable  
representations for the bound state vertices for arbitrary quantum numbers $J^{PC}$. 

As an illustration of the framework, we have calculated the scalar and 
pseudoscalar glueball mass. Our result for the scalar glueball state is 
certainly encouraging, although in the pseudoscalar channel the mass is 
exceedingly high. Compared to the recent exploratory approach of 
Ref.~\cite{Meyers:2012ka} we
have made a number of technical improvements. Most important are the use of
explicit solutions of the ghost and gluon propagators in the complex momentum
plane. Furthermore, our approach fully maintains multiplicative renormalizability.

Our framework for calculations of glueball properties offers various prospects of 
improvements and applications in the near future. First, different ansaetze 
for the three-gluon vertex should be used and its impact on the spectrum 
analyzed. It would be desirable, although technically very demanding, to use dynamical 
three-point vertices as in \cite{Huber:2012kd} and to include the four-gluon interaction 
contributions \cite{Binosi:2014kka,Cyrol:2014kca} into our framework.
On the other hand, a very important extension is the inclusion of meson/glueball mixing 
along the lines of Eqs.~\eqref{eq:QCDBSE1} to \eqref{eq:QCDBSE3}.
This will allow to leave the sector of pure gauge field calculations 
of glueballs and thus provide access to realistic glueball properties 
in the future.

\vspace{.2cm}

\noindent{\bf Acknowledgements}\\
We thank Reinhard Alkofer and Richard Williams for fruitful 
discussions. This work was supported by an 
Erwin Schr\"odinger fellowship J3392-N20 
from the Austrian Science Fund (FWF), the Helmholtz Young Investigator 
Grant VH-NG-332, the Helmholtz International Center for FAIR within the LOEWE 
program of the State of Hesse, the European Commission
FP7-PEOPLE-2009-RG No.~249203, and by the BMBF contract 06GI7121.

\end{document}